\documentstyle[prl,floats,twocolumn,aps,epsf]{revtex}

\begin{document}
\draft
\title{Dense coding for continuous variables}
\author{Samuel L.~Braunstein}
\address{SEECS, University of Wales, Bangor LL57 1UT, UK}
\author{H.\ J.\ Kimble}
\address{Norman Bridge Laboratory of Physics 12-33, \\
California Institute of Technology, Pasadena, CA 91125}
\date{\today}
\maketitle

\begin{abstract}
A scheme to achieve dense quantum coding for the quadrature amplitudes of
the electromagnetic field is presented. The protocol utilizes shared
entanglement provided by nondegenerate parametric down conversion in the
limit of large gain to attain high efficiency. For a constraint in the mean
number of photons $\bar n$ associated with modulation in the signal channel,
the channel capacity for dense coding is found to be $\ln(1+\bar n+\bar n^2)$%
, which always beats coherent-state communication and surpasses
squeezed-state communication for $\bar n > 1$. For $\bar n\gg 1$, the dense
coding capacity approaches twice that of either scheme.
\end{abstract}



An important component of contemporary quantum information theory is the
investigation of the classical information capacities of noisy quantum
communication channels. Here, classical information is encoded by the choice
of one particular quantum state from among a predefined ensemble of quantum
states by the sender {\it Alice\/} for transmission over a quantum channel
to the receiver {\it Bob}. If Alice and Bob are allowed to communicate only
via a one-way exchange along such a noisy quantum channel, then the optimal
amount of classical information that can be reliably transmitted over the
channel has recently been established \cite{Hole96,Schu97}.

Stated more explicitly, if a classical signal $\alpha $ taken from the
ensemble $P_{\alpha }$ is to be transmitted as a quantum state $\hat\rho%
_\alpha$, then Holevo's bound for a bosonic quantum channel says that the
mutual information $H(A:B)$ between the sender $A$ ({\it Alice}) and
receiver $B$ ({\it Bob}) is bounded by \cite{Hole96} 
\begin{equation}
H(A:B)\leq S(\hat\rho )-\int d^{2}\alpha \,P_{\alpha }S(\hat\rho_\alpha)
\leq S(\hat\rho ) \;,  \label{Holevo}
\end{equation}
where $S(\hat\rho)$ is the von Neumann entropy associated with the density
operator $\hat\rho =\int d^{2}\alpha\,P_\alpha\,\hat\rho_\alpha$ for the
mean channel state.

By contrast, if Alice and Bob share a quantum resource in the form if an
ensemble of entangled states, then quantum mechanics enables protocols for
communication that can circumvent the aforementioned bound on channel
capacity. For example, as shown originally by Bennett and Wiesner \cite
{densecoding}, Alice and Bob can beat the Holevo limit by exploiting their
shared entanglement to achieve dense quantum coding. Here, the signal is
encoded at {\it Alice's sending station\/} and transmitted via one component
of a pair of entangled quantum states, with then the second component of the
entangled pair exploited for decoding the signal at {\it Bob's receiving
station}. In this scheme, the cost of distributing the entangled states to 
{\it Alice\/} and {\it Bob\/} is not figured into the accounting of
constraints on the quantum channel (e.g., the mean energy). Such neglect of
the distribution cost of entanglement is sensible in some situations, as for
example, if the entanglement were to be sent during off-peak times when the
communication channel is otherwise under utilized, or if it had been
conveyed by other means to Alice and Bob in advance (e.g., via a pair of 
{\it quantum CDs\/} with stored, entangled quantum states). Note that in
general, no signal modulation is applied to the second (i.e., {\it Bob's})
component of the entangled state, so that it carries no information by
itself.

Although quantum dense coding has most often been discussed within the
setting of {\it discrete\/} quantum variables (e.g., {\it qubits}) \cite
{densecoding,Mattle}, in this paper we show that highly efficient dense
coding is possible for {\it continuous\/} quantum variables. As in our prior
work on quantum teleportation \cite{sam98a,sam98b,akira98}, our scheme for
achieving quantum dense coding exploits squeezed-state entanglement, and
therefore should allow {\it unconditional\/} signal transmission with high
efficiency, in contrast to the {\it conditional} transmission with extremely
low efficiency achieved in Ref. \cite{Mattle}. More specifically, for signal
states $\alpha $ associated with the complex amplitude of the
electromagnetic field, the channel capacity for dense coding is found to be $%
\ln (1+\bar{n}+\bar{n}^{2})$, where $\bar{n}$ is the mean photon number for
modulation in the signal channel. The channel capacity for dense coding in
our scheme thus always beats coherent-state communication and surpasses
squeezed-state communication for $\bar{n}>1$. For $\bar{n}\gg 1$, the dense
coding capacity approaches twice that of either scheme.

As illustrated in Fig.~\ref{fig1}, the relevant continuous variables for our
protocol are the quadrature amplitudes $(\hat{x},\hat{p})$ of the
electromagnetic field, with the classical signal $\alpha =\langle \hat{x}%
\rangle +i\langle \hat{p}\rangle $ then associated with the quantum state $%
\hat{\rho}_{\alpha }$ drawn from the phase space for a single mode of the
field. The entangled resource shared by {\it Alice\/} and {\it Bob\/} is a
pair of EPR beams with quantum correlations between canonically conjugate
variables $(\hat{x},\hat{p})_{(1,2)}$ as were first described by Einstein,
Podolsky, and Rosen (EPR \cite{EPR}), and which can be efficiently generated
via the nonlinear optical process of parametric down conversion, resulting
in a highly squeezed two-mode state of the electromagnetic field \cite
{Reid,Ou92}. In the ideal case, the correlations between quadrature-phase
amplitudes for the two beams $(1,2)$ are such that 
\begin{equation}
\langle (\hat{x}_{1}-\hat{x}_{2})^{2}\rangle \rightarrow 0\;, ~~~ \langle (%
\hat{p}_{1}+\hat{p}_{2})^{2}\rangle \rightarrow 0\;,  \label{delta-xp}
\end{equation}
albeit it with an concomitant divergence in the mean photon number $\bar{n}$
in each channel.

Component $1$ of this entangled pair of beams is input to {\it Alice's
sending station}, where the message $M_{a}^{\alpha }$ corresponding to the
classical signal $\alpha _{{\rm in}}$ is encoded as the quantum state $\hat{%
\rho}_{\alpha _{{\rm in}}}$ by a simple phase-space offset by way of the
displacement operator $\hat{D}(\alpha_{{\rm in}})$ applied to $1$ \cite
{mandel-wolf}. The displacement $\hat{D}(\alpha _{{\rm in}})$ can be
implemented in a straightforward fashion by amplitude and phase offsets
generated by the (suitably normalized) classical currents $%
(i_{x_{a}},i_{p_{a}})$ as in Ref.~\onlinecite{akira98}.  The state
corresponding to Alice's displacement of the EPR beam constitutes the
quantum signal and is transmitted along the quantum channel shown in Fig.~%
\ref{fig1} to {\it Bob's receiving station}, where it is decoded with the
aid of the second component $2$ of the original EPR pair of beams and the
homodyne detectors $(d_{x},d_{p})$. The resulting photocurrents $%
(i_{x_{b}},i_{p_{b}})$ suitably normalized to produce $\alpha_{{\rm out}%
}=i_{x_{b}}+i\,i_{p_{b}}$ constitute the message $M_{b}^{\alpha }$ received
by Bob. In the limit $\bar{n}\rightarrow \infty $, Eq.~(\ref{delta-xp})
ensures $\alpha _{{\rm out}}=\alpha _{{\rm in}}$, so that the classical
message would be perfectly recovered. However, even for finite $\bar{n}$ as
is relevant to a channel constrained in mean energy, the finite correlations
implicit in the EPR beams enable quantum dense coding with enhanced channel
capacity relative to either coherent state or squeezed state communication,
as we now show.

Consider the specific case of EPR beams $(1,2)$ approximated by the two-mode
squeezed state with Wigner function 
\begin{eqnarray}
W_{{\rm EPR}}&&(\alpha_{1},\alpha_{2}) \\
&&={\frac{4}{\pi ^{2}}}\exp [-e^{-2r}(\alpha
_{1}-\alpha_{2})_{R}^{2}-e^{2r}(\alpha_{1}-\alpha_{2})_{I}^{2}  \nonumber \\
&&\phantom{={\frac{4}{\pi ^{2}}}\exp} -e^{2r}(\alpha_{1}+%
\alpha_{2})_{R}^{2}-e^{-2r}(\alpha _{1}+\alpha _{2})_{I}^{2}] \;,   \nonumber
\end{eqnarray}
where the subscripts $R$ and $I$ refer to real and imaginary parts of the
field amplitude $\alpha $, respectively (i.e., $\alpha _{R,I}=x,p$). Note
that for $r\rightarrow \infty $, the field state becomes the ideal EPR state
as described in Eq.~(\ref{delta-xp}), namely 
\begin{equation}
W_{{\rm EPR}}(\alpha _{1},\alpha _{2})\rightarrow C\,\delta (\alpha
_{1R}+\alpha _{2R})\,\delta (\alpha _{1I}-\alpha _{2I})\;.
\end{equation}

As shown in Fig.~\ref{fig1}, signal modulation is performed only on mode $1$%
, with mode $2$ treated as an overall shared resource by {\it Alice\/} and 
{\it Bob\/} (and which could have been generated by {\it Alice\/} herself).
The modulation scheme that we choose is simply to displace mode $1$ by an
amount $\alpha _{{\rm in}}$. This leads to a displaced Wigner function given
by $W_{{\rm EPR}}(\alpha _{1}-\alpha _{{\rm in}},\alpha _{2})$,
corresponding to the field state that is sent via the quantum channel from 
{\it Alice\/} to {\it Bob}.

Upon receiving this transmitted state (consisting of the modulated mode $1$%
), the final step in the dense-coding protocol is for {\it Bob\/} to combine
it with the shared resource (mode $2$) and retrieve the original classical
signal $\alpha _{{\rm in}}$ with as high a fidelity as possible. As
indicated in Fig.~\ref{fig1}, this demodulation can be performed with a
simple $50-50$ beam splitter that superposes the modes $(1,2)$ to yield
output fields that are the sum and difference of the input fields and which
we label as $\beta _{1}$ and $\beta _{2}$, respectively. The resulting state
emerging from {\it Bob's\/} beamsplitter has Wigner function 
\begin{eqnarray}
&&W_{{\rm sum/diff}}(\beta _{1},\beta _{2}) \\
&&~~=W_{{\rm EPR}}\biglb((\beta _{1}+\beta _{2})/\sqrt{2}-\alpha ,(\beta
_{1}-\beta _{2})/\sqrt{2}\bigrb)\;.  \nonumber
\end{eqnarray}
The classical signal that we seek is retrieved by homodyne detection at
detectors $(d_{x},d_{p})$, which measure the analogs of position and
momentum for the sum and difference fields $(\beta _{1},\beta _{2})$. For
ideal homodyne detection the resulting outcomes are distributed according to 
\[
P(\beta |\alpha )={\frac{2e^{2r}}{\pi }}\exp (-2e^{2r}|\beta -\alpha /\sqrt{2%
}|^{2})\;,
\]
where $\beta =\beta _{1R}+i\beta _{2I}$ and represents a highly peaked
distribution about the complex displacement $\alpha /\sqrt{2}$. For large
squeezing parameter $r$ this allows us to extract the original signal $%
\alpha $ which we choose to be distributed as 
\begin{equation}
P_{\alpha }={\frac{1}{\pi \sigma ^{2}}}\exp (-|\alpha |^{2}/\sigma ^{2})\;.
\end{equation}
Note that mode 1 of this displaced state has a mean number of photons given
by 
\begin{equation}
\bar{n}=\sigma ^{2}+\sinh ^{2}r\;.  \label{nbar}
\end{equation}

In order to compute the quantity of information that may be sent through
this dense coding channel we note the unconditioned probability for the
homodyne statistics is given by 
\begin{equation}
P(\beta)={\frac{2}{\pi(\sigma^2 +e^{-2r})}} \exp\left(\frac{-2|\beta|^2}{%
\sigma^2 +e^{-2r}}\right) \;.
\end{equation}
The mutual information describing the achievable information throughput of
this dense coding channel is then given by 
\begin{eqnarray}
H^{{\rm dense}}(A:B) &=& \int d^2\beta\, d^2\alpha\, P(\beta|\alpha)P_\alpha
\ln\left( {\frac{P(\beta|\alpha)}{P(\beta)}}\right)  \nonumber \\
&=& \ln(1+\sigma^2e^{2r}) \;.
\end{eqnarray}
For a fixed $\bar n$ in Eq.~(\ref{nbar}) this information is optimized when $%
\bar n = e^r\sinh r$, i.e., when $\sigma^2 = \sinh r\cosh r$ so yielding a
dense coding capacity of 
\begin{equation}
C^{{\rm dense}} = \ln (1+\bar n+\bar n^2) \;,  \label{dc}
\end{equation}
which for large squeezing $r$ becomes 
\begin{equation}
C^{{\rm dense}} \sim 4 r \;.  \label{dclarger}
\end{equation}

How efficient is this dense coding in comparison to single channel coding?
Let us place a `common' constraint of having a fixed mean number of photons $%
\bar n$ which can be modulated. For a single bosonic channel Drummond and
Caves \cite{Drummond} and Yuen and Ozawa \cite{Yuen} have used Holevo's
result to show that the optimal channel capacity is just that given by
photon counting from a maximum entropy ensemble of number states. In this
case the channel capacity (the maximal mutual information) achieves the
ensemble entropy, see Eq.~(\ref{Holevo}), so 
\begin{equation}
C=S(\rho) = (1+\bar n)\ln (1+\bar n)-\bar n\ln \bar n \;.
\end{equation}
Substituting $\bar n = e^r \sinh r$ into this we find 
\begin{equation}
C\sim 2 r \;,
\end{equation}
for large squeezing $r$. This is just one-half of the asymptotic dense
coding mutual information, see Eq.~(\ref{dclarger}). Thus asymptotically, at
least, the dense coding scheme allows twice as much information to be
encoded within a given state, although it has an extra expense (not included
within the simple constraint $\bar n$) of requiring shared entanglement.

It is worth noting that this dense coding scheme does {\it not\/} always
beat the optimal single channel capacity. Indeed, for small squeezing it is
worse. The break-even squeezing required for dense coding to equal the
capacity of the optimal single channel communication is 
\begin{equation}
r_{{\rm break-even}} \simeq 0.7809 \;,
\end{equation}
which corresponds to roughly $6.78\, dB$ of two-mode squeezing or to $\bar n
\simeq 1.884$. This break-even point takes into account the difficulty of
making highly squeezed two-mode squeezed states. No similar difficulty has
been factored into making ideal number states used in the benchmark scheme
with which our dense coding scheme is compared.

A fairer comparison is against single-mode coherent state communication with
heterodyne detection. Here the channel capacity is well known \cite
{Gordon,She,Yam} for the mean photon number constraint to be 
\begin{equation}
C^{{\rm coh}} = \ln(1+\bar n) \;,
\end{equation}
which is {\it always\/} beaten by the optimal dense coding scheme described
by Eq.~(\ref{dc}).

An improvement on coherent state communication is squeezed state
communication with a single mode. The channel capacity of this channel has
been calculated \cite{Yam} to be 
\begin{equation}
C^{{\rm sq}} = \ln(1+2\bar n) \;,
\end{equation}
which is beaten by the dense coding scheme of Eq.~(\ref{dc}) for $\bar n > 1$%
, i.e., the break-even squeezing required is 
\begin{equation}
r^{{\rm sq}}_{{\rm break-even}} \simeq 0.5493 \;,
\end{equation}
which corresponds to $4.77\, dB$.

In summary, we have shown how to perform dense quantum coding for continuous
quantum variables by utilizing squeezed state entanglement. For a constraint
in the mean number of photons that may be modulated $\bar{n},$ the dense
coding capacity is found to be $\ln (1+\bar{n}+\bar{n}^{2})$. This scheme
always beats single mode coherent state communication and surpasses single
mode squeezed state communication for $\bar{n}>1$. Note that in terms of
actual implementation, our protocol should allow for high efficiency, {\it %
unconditional\/} transmission with encoded information sent every inverse
bandwidth time. This situation is in contrast to implementations that employ
weak parametric down conversion, where transmission is achieved {\it %
conditionally\/} and relatively rarely. In fact Mattle et al.\ \cite{Mattle}
obtained rates of only 1 in $10^{7}$ per inverse bandwidth time \cite{Wein}.
By going to strong down conversion and using a characteristically different
type of entanglement, our scheme should allow information to be sent with
much higher efficiency and should simultaneously improve the ability to
detect orthogonal Bell states. Indeed, these advantages enabled the first
experimental realization of unconditional quantum teleportation within the
past year \cite{akira98}. Beyond the particular setting of quantum
communication discussed here, this research is part of a larger program to
explore the potential for quantum information processing with continuous
quantum variables. Such investigations are quite timely in light of
important recent progress concerning the prospects for diverse quantum
algorithms with continuous variables, including universal quantum
computation \cite{lloyd98a} and quantum error correction \cite
{lloyd98b,sam98i,sam98ii}, with quantum teleportation being a prime example 
\cite{sam98a,ben93,vaidman}. Although still in its earliest stages,
theoretical protocols have been developed for realistic physical systems
that should allow a variety of elementary processing operations for
continuous quantum variables, including significantly quantum storage for
EPR\ states.\cite{scott99a,scott99b}

\vskip 0.1truein

SLB was supported in part by the UK Engineering and Physical Sciences
Research Council and the Royal Academy of Engineering. The work of HJK is
supported by DARPA via the QUIC Institute which is administered by ARO, by
the Office of Naval Research, and by the National Science Foundation.

\begin{figure}[thb]
\caption{Illustration of the scheme for achieving super-dense quantum coding
for signal states over the complex amplitude $\protect\alpha =x+ip$ of the
electromagnetic field. The quantum resource that enables dense coding is the
EPR source that generates entangled beams $(1,2)$ shared by {\it Alice\/}
and {\it Bob}.}
\label{fig1}
\end{figure}

\begin{figure}[thb]
\caption{Depiction of signal decoding at {\it Bob's receiving station}. (a)
At Bob's $50-50$ beam splitter $m_{b}$, the displaced EPR beam $1$ is
combined with the component $2$ to yield two independent squeezed beams,
with the $\protect\beta _{1,2}$ beams having fluctuations reduced below the
vacuum-state limit along $(x_{\protect\beta _{1}},p_{\protect\beta _{2}})$.
Homodyne detection at $(d_{x},d_{p})$ (Fig.~\ref{fig1}) with LO phases set
to measure $(x_{\protect\beta _{1}},p_{\protect\beta _{2}})$, respectively,
then yields the complex signal amplitude $\protect\alpha _{{\rm out}}$ with
variance set by the associated squeezed states. (b) The net effect of the
dense coding protocol is the transmission and detection of states of complex
amplitude $\protect\alpha $ with an effective uncertainty below the
vacuum-state limit (indicated by the dashed circle).}
\label{fig2}
\end{figure}

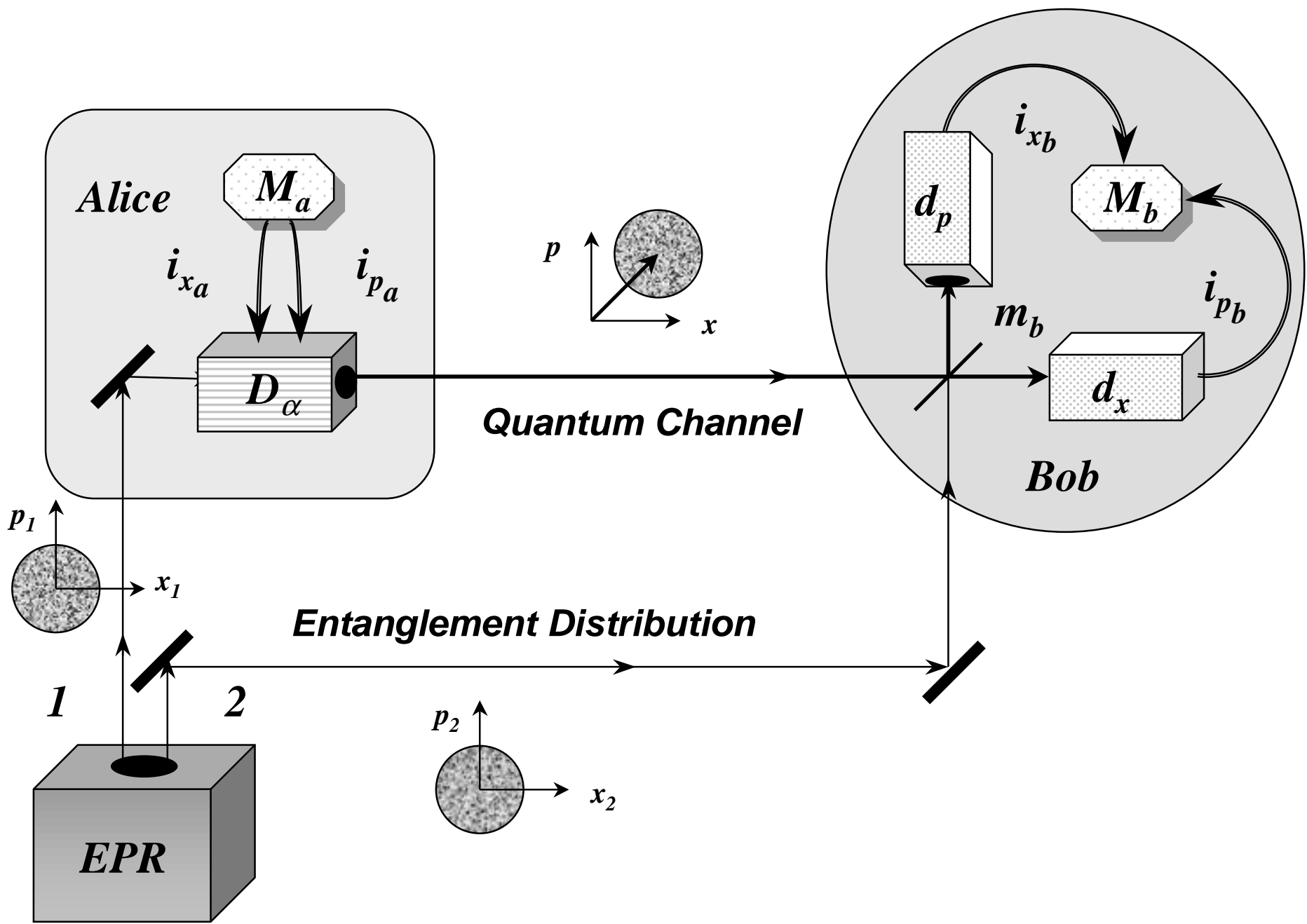
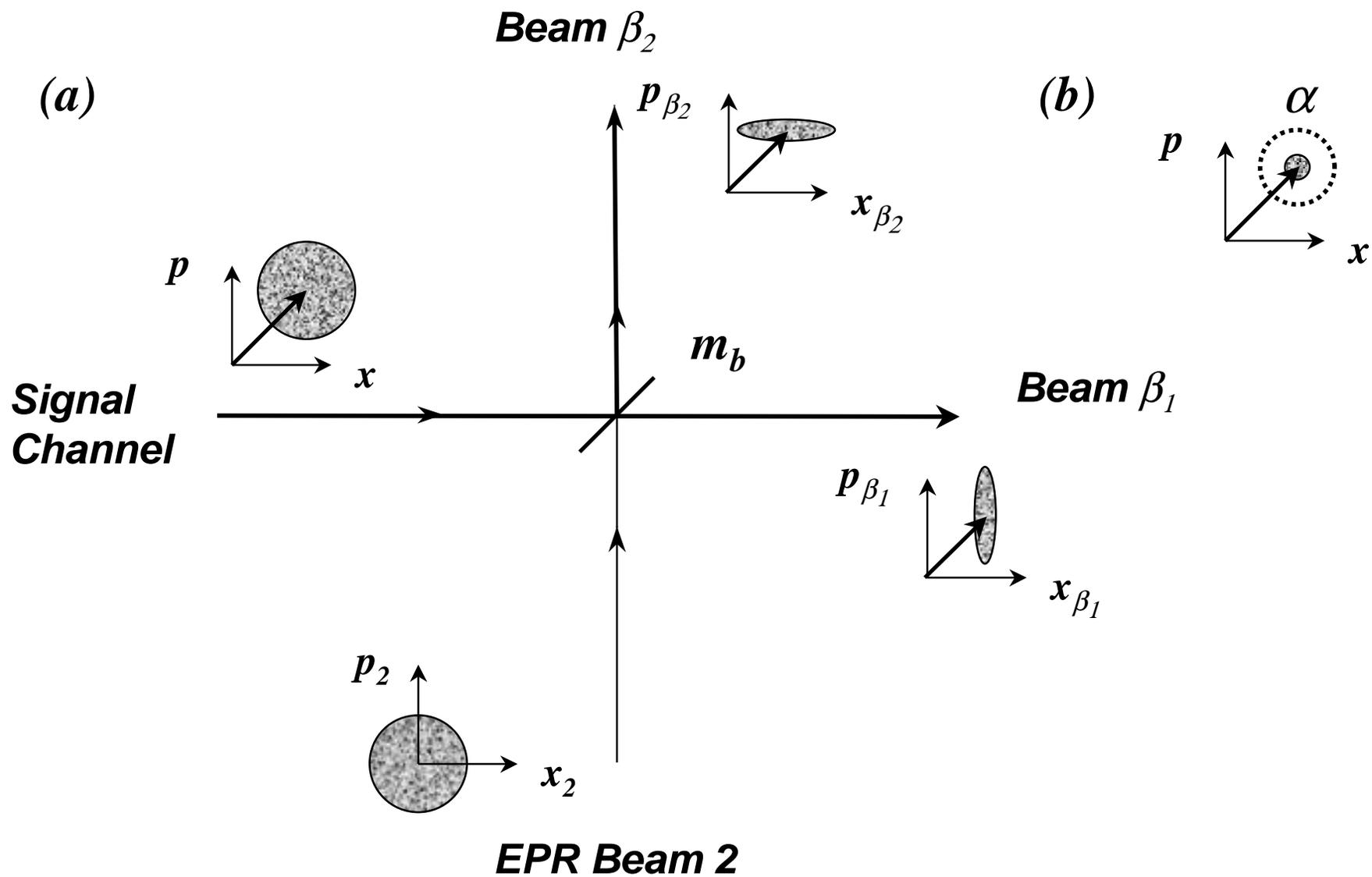
\end{document}